\documentclass[12pt]{article}
\usepackage{graphicx,amsmath}
\usepackage{units}
\usepackage{color}

\newlength{\dinwidth}
\newlength{\dinmargin}
\def\lapproxeq{\lower .7ex\hbox{$\;\stackrel{\textstyle                                                    
<}{\sim}\;$}}                                                    
\def\gapproxeq{\lower .7ex\hbox{$\;\stackrel{\textstyle                                                    
>}{\sim}\;$}}                                                    
\def\be{\begin{equation}}                                                    
\def\ee{\end{equation}}                                                    
\def\bea{\begin{eqnarray}}                                                    
\def\eea{\end{eqnarray}}

\def\sh{\hat s}
\def\sh2{{\hat s}^2}

\begin{document}

\begin{flushright}                                                    
IPPP/14/54  \\
DCPT/14/108 \\                                                    
\today \\                                                    
\end{flushright} 

\vspace*{0.5cm}

\begin{center}
{\Large \bf BFKL equation for an integrated gluon density}\\

\vspace*{1cm}
                                                   
E.G. de Oliveira$^a$, A.D. Martin$^b$ and M.G. Ryskin$^{b,c}$  \\                                                    
                                                   
\vspace*{0.5cm}                                                    
$^a$ Departamento de F\'{i}sica, CFM, Universidade Federal de Santa Catarina, 
C.P. 476, CEP 88.040-900, Florian\'opolis, SC, Brazil \\
$^b$ Institute for Particle Physics Phenomenology, University of Durham, Durham, DH1 3LE \\                                                   
$^c$ Petersburg Nuclear Physics Institute, NRC Kurchatov Institute, Gatchina, St.~Petersburg, 188300, Russia \\          
                                                    
\vspace*{1cm}

\begin{abstract} 

We show how it is possible to rewrite the BFKL equation for the unintegrated  gluon distribution, in terms of integrated gluons, similar to that used in DGLAP. We add to our equation the next-to-leading log terms which provide exact energy-momentum conservation and account for the kinematic constraint in real gluon emission. In this way the equation includes the major part of the higher-order corrections to BFKL evolution. We discuss the possibility to obtain a unified BFKL-DGLAP evolution equation relevant to processes at the LHC where both log$(1/x)$ and log$Q^2$ are large simultaneously.

\end{abstract}

\end{center}

\section{Introduction  \label{sec:1}} 
Originally the BFKL equation~\cite{BFKL} was proposed to describe the high-energy behaviour of processes involving hadrons, such as proton-proton scattering or deep inelastic lepton-proton scattering. The BFKL equation for the amplitude of such processes sums up all the higher order $\alpha_s$ corrections
where the small value of QCD coupling $\alpha_s$ is compensated by the large logarithm of the energy, $\sqrt s$; that is the amplitude has the form 
\be
A(s)~=~\sum_n C_n(\alpha_s\ln s)^n\ .
\label{eq:LL}
\ee
Recalling that $x \sim Q^2/s$, where $Q^2$ is the hard scale of the process, it follows that the BFKL equation may be regarded as an  equation for the  $\ln(1/x)$ evolution
of {\it unintegrated} gluon density of the proton, $f(x,k_t)$, which depends on two arguments: the proton momentum fraction $x$ carried by a gluon and its transverse momentum, $k_t$. A feature of this evolution is diffusion of the gluon density in ln$k_t$ space.  

On the other hand, DGLAP evolution, which sums the $\alpha_s{\rm ln}Q^2$ terms, has contributions which are strongly ordered in $k_t$, and is written in terms of {\it integrated} parton densities which no longer depend on $k_t$.  The domains in the $(x,Q^2)$ plot, where pure BFKL and pure DGLAP evolution are appropriate, are quite distinct.  In practice, however, we really should sum up both the BFKL and the DGLAP logarithms. In particular, at the LHC energy of $\sqrt s=14$ TeV the most interesting kinematical domain corresponds to the scale $Q^2\sim M_W^2\sim 6400$ GeV$^2$ (that is $\ln(Q^2/Q^2_0)\sim 8$) and $x\sim M_W/\sqrt s\sim 0.005$ (that is $\ln(1/x)\sim 5$).
It is a region where both the BFKL and the DGLAP logarithms are important.

In order to be able to compare the BFKL and DGLAP evolutions  
it would be valuable to express the BFKL equation in terms of an {\it integrated} gluon 
distribution, as conventionally used in the DGLAP approach.  Moreover, this would open the way to formulate an expression which accounts for both the DGLAP and the BFKL logarithms in terms of integrated densities.\footnote{Such a unified BFKL--DGLAP equation was proposed in~\cite{KMS}, but in terms of the {\em unintegrated} parton densities. Note that it is not the same as the resummation of the large, enhanced by $\ln(1/x)$, BFKL contributions to the DGLAP splitting functions, that is to the anomalous dimensions, as was done in \cite{ABF}. Recall that, besides the leading twist, the BFKL
equation includes higher-twist effects which account for gluon reggeization.} 

 In this form, it would be easier to study the BFKL effects, caused by contributions which violate the strong-$k_t$ ordering or by higher-twist contributions (both of which are present in BFKL, but are absent in DGLAP evolution), and/or to perform a BFKL-based global analysis
analogous to the traditional global parton analyses made within the DGLAP approach. 

Recall that the BFKL equation describes evolution in the $\ln(1/x)$ direction starting from some input (which depends on $Q^2$) at fixed $x=x_0$, while DGLAP
generates $\ln Q^2$ evolution starting from input at fixed $Q^2=Q^2_0$. 
The power of the $x$-dependence in DGLAP evolution is driven mainly by the input distribution.
On the other hand, in the BFKL approach the small $x$ behaviour is completely determined by the BFKL equation. 

To discover such a BFKL equation for an `integrated' gluon density is the purpose of this paper~\footnote{At first sight, such a BFKL equation for an `integrated' gluon density was already presented long ago in~\cite{BF}. However the equation proposed in~\cite{BF} does not account properly for gluon reggeization and for the running of the QCD coupling $\alpha_s$. In Section 5 we will discuss these problems in more detail. }.   Can this be done, so that the low $x$ dependence of the integrated gluon PDF is completely generated within the BFKL framework? It will clearly involve higher-twist effects coming from the reggeization of the gluon, and hence lie outside a pure DGLAP framework (which is based on leading twist only). However, if it can be done, then it will open the way to obtaining a DGLAP-like evolution for an integrated gluon PDF in terms of a single evolution variable which sums both the BFKL log$(1/x)$ and DGLAP log$Q^2$ contributions. We comment further on this attractive possibility in Section 5.

Note that, by summing up all the $\ln(1/x)$-enhanced contributions, the BFKL equation deals with kinematics where the fraction, $z$, of the parent gluon momentum carried by the following gluon is small; $z\ll1$. In this situation one may neglect the momenta of the new gluons in comparison with the momentum of parent gluon. However, in reality, the typical values of $z$ are not so small. Therefore 
the parton distribution, generated by BFKL evolution, violates the energy-momentum conservation law. Formally this violation is a next-to-leading Log (NLL) effect, but numerically it may be important. Moreover, since the majority of available data comes from Deep Inelastic Scattering where the incoming photon does not interact with the gluon directly, in the global parton analyses the normalisation of gluon distribution
 is mainly fixed by the energy conservation sum rule. Therefore it is crucial to have an equation which automatically satisfies the energy-momentum conservation law.
For this reason, in the equation for integrated gluon density that we propose, we include the NLL term which restores energy-momentum conservation.  

Besides this, we will take care of the kinematical cutoff $k^{'2}_t<k^2_t/z$~\cite{Ci,KMS1,Bo} for real gluon emission. Again, formally, since $z\ll1$, the integral over the intermediate momentum $k'_t$ may run up to $k'_t\to\infty$. However, in spite of the fact that this integral is well convergent at $k'_t\gg k_t$, for too large $k^{'2}_t>k^2_t/z$ the contribution of the longitudinal component to the virtuality $k^{'2}$ becomes so large that it kills the leading logarithmic form of $dz/z$ integration, providing, in this way, an effective cutoff $k^{'2}_t<k^2_t/z$. 

The inclusion of this cutoff (sometimes called the `consistency constraint') explicitly in the BFKL equation was found to play a crucial role; it accounts for a major part of the NLL and higher-order corrections.  Let us explain the importance of this statement. The BFKL resummation of the ln$(1/x)$ contributions shown in (\ref{eq:LL}) is at Leading Log (LL) level. It results in a gluon density which behaves as $f \propto x^{-\omega_0}$ as $x\to 0$, where $\omega_0={\bar \alpha}_s4{\rm ln}2$, with ${\bar \alpha}_s\equiv 3\alpha_s/\pi$.  Resumming the next-to-leading logs \cite{FLnll} gives a behaviour of the form $f \propto x^{-\omega}$ where now 
\be
\omega=\omega_0(1-6.5{\bar \alpha}_s).
\ee
At first it was thought that such a large NLL correction would mean that no stable small $x$ predictions could be made using the BFKL procedure. Next, the value of the NLL correction depends on the  choice of renormalization scheme. Recently, it was shown however~\cite{CWS} that applying the principle of maximum conformality and using the more physical MOM-scheme a much more stable value of $\omega$ can be obtained. Moreover, 
  it is possible to identify higher-order terms and then to resum them. Indeed Ciafaloni et al. \cite{CCS} carried out an all-order ln$(1/x)$ resummation of the following effects: (i) running $\alpha_s$, (ii) the non-singular DGLAP terms and (ii) the angular ordering and energy constraints. The result was a stable $x^{-\omega}$ behaviour, which is consistent with observations. In fact, prior to this, the fit in \cite{KMS}, which was based on a unified BFKL/DGLAP equation for the unintegrated gluon,  incorporates these all-order ln$(1/x)$ contributions, where the imposition of the kinematic (or so-called consistency) constraint \cite{KMS1} plays a major role.

In Section 2 we recall the original BFKL equation for the {\em unintegrated} gluon density, and express it in a form already including the kinematical cutoff, $k^{'2}_t<k^2_t/z$. Also, here, we add the next-to-leading term which restores energy-momentum conservation. Then, in Section 3, we introduce the {\em integrated} gluon distribution $F(x,q)$ and obtain the equation for $F$. The final expression is given in Section 4, and an overview presented in Section 5.

\section{BFKL equation for unintegrated gluons}

Recall that the BFKL equation can be written as an integral equation for the unintegrated gluon distribution $f(x,k_t)$ (as conventionally used in BFKL evolution) in the form:
\be
\label{eq:uBFKL}
f(x,k_t)=f_0(x,k_t) + \frac{\alpha_s}{2\pi}  \int_x^1 \frac{d z}{z} \int_{k_0}^\infty \frac{d^2k'_t}{\pi} ~{\cal K}(k_t,k'_t,z)~f(x/z,k'_t),
\ee
where the kernel is evaluated as
\be
\label{eq:kernel}
{\cal K}(k_t,k_t',z)f(x/z,k'_t)=
2N_c\frac{k^2_t}{k^{'2}_t}\left[\frac{\Theta(k^2_t/z-k^{'2}_t)f(x/z,k'_t)-f(x/z,k_t)}{|k^{'2}_t-k^2_t|}
~+~
\frac{f(x/z,k_t)}{\sqrt{4k^{'4}_t+k^4_t}}\right]~.
\ee
The first term with the $\Theta$ function in the kernel\footnote{Here we have already integrated over the azimuthal angle $\phi$ assuming, similar to DGLAP case, a flat $\phi$ dependence of $f$; that is, we consider the zero harmonic, which corresponds to the rightmost intercept.} can be understood as the effect of the emission of a daughter gluon with momentum ($x,k_t$) from a parent gluon with momentum ($x'=x/z, k_t'$). The remaining terms (with $f(x/z,k_t)$) accounts for the loop correction originated by the trajectory of $t$-channel reggeized gluons. Note that now the BFKL kernel ${\cal K}$ contains a third argument $z$ since here we have implemented the important kinematic constraint for real emission \cite{Ci,KMS1,Bo,KMS} 
\be
k_t^{'2}<\frac{k^2_t}z  ~,
\label{eq:constraint}
\ee
to guarantee that only the leading logarithm part of the higher-order corrections is actually taken into account.

 To ensure energy conservation, we subtract the term
\be
 \frac{\alpha_s}{2\pi} \int_0^1dz \int_{k_o}^\infty \frac{d^2k'_t}{\pi}~{\cal K}
(k_t,k'_t,z)f(x,k'_t) 
\label{eq:substr}
\ee
from the right-hand side of (\ref{eq:uBFKL}).
We can check that this conserves momentum by integrating both (\ref{eq:uBFKL}) and (\ref{eq:substr}) with respect to $x$, and using the ``integral'' identity
\be
 \int_0^1 dx  \int_x^1 \frac{d z}{z} =  \int_0^1 \frac{d z}{z}  \int_0^z dx
  = \int_0^1 d z \int_0^1 d \left( \frac{x}{z} \right) 
\ee

Such a prescription is analogous to the $1/\omega\to 1/\omega-1$ replacement proposed in~\cite{EL} to achieve the same goal in $\omega$-representation.  
Note that the integral over $z$ in (\ref{eq:substr}) does not have a logarithmic ($dz/z$) form, whereas in (\ref{eq:uBFKL}) the form $(dz/z)$ generates the factor $1/\omega$ in $\omega$ representation. Recall that $f(x,k_t)\propto x^{-\omega}$.


\section{BFKL equation for integrated gluons}

In this section we show how the BFKL equation can be rewritten in terms of the {\em integrated} gluon distribution given by
\be
\label{eq:F}
F(x,q)=\int^{q^2} \frac{d k_t^2}{k_t^2} \, f(x,k_t)
=\int^{q^2} d \ln k_t^2 \, f(x,k_t).
\ee 
First, we integrate (\ref{eq:uBFKL}) over $dk_t^2/k_t^2$ from $k_0^2$ to $q^2$, and express it in the form
\be
\label{eq:ABC}
F(x,q) - F(x,k_0) = F_0(x, q) - F_0(x,k_0) 
+ \frac{N_c \alpha_s}{\pi} \int_x^1 \frac{d z}{z} \left( F_A + F_B + F_C \right),
\ee
where $F_A$, $F_B$, and $F_C$ arise from the three terms in [...] of (\ref{eq:kernel}), but now integrated over both $k_t$ and $k'_t$; and where
\be
F_0(x,q)=\int^{q^2} \frac{dk^2_t}{k^2_t}f_0(x,k_t).
\ee 

We first consider $F_C$, which is given by
\bea
F_C & = & \int_{k_0^2}^{q^2} \frac{d k_t^2}{k_t^2}
 \int_{k_0^2}^\infty d {k'_t}^2
 \frac{k^2_t}{k^{'2}_t}\left[
\frac{f(x',k_t)}{\sqrt{4k^{'4}_t+k^4_t}} \right] \\
 & = &  \int_{k_0^2}^{q^2} \frac{d k_t^2}{k_t^2} \left[ \ln \left( \frac{\sqrt{4 k_0^4+k^4_t} + k^2_t}{2 k_0^2} \right) f(x',k_t) \right].  \label{eq:11}
\eea
We have carried out one of the integrations, but so far our goal of having only an integrated distribution has not been achieved. To do so, we perform an `integration by parts' using the identity
\begin{equation}
\int u d v = \int d( u v) - \int v du.
\end{equation}
Note that on the right-hand side we only have an integrated $v$,  whereas on the left-hand we have an unintegrated $dv$. That is exactly what we need.  In particular, to evaluate (\ref{eq:11}) we use
\be
u  \equiv  \ln \left( \frac{\sqrt{4 k_0^4+k^4_t} + k^2_t}{2 k_0^2} \right) ~~~~~~~~{\rm and}~~~~~~~~
d v  \equiv   \frac{d k_t^2}{k_t^2} f(x',k_t)
\ee
and integrate by parts. We obtain
\bea \nonumber
F_C & = & \left[ \ln \left( \frac{\sqrt{4 k_0^4+k^4_t} + k^2_t}{2 k_0^2} \right) F(x',k_t) \right] \Bigg|_{k_t^2=k_0^2}^{k_t^2=q^2} 
 -   \int_{k_0^2}^{q^2} d k_t^2 \frac{F(x',k_t)}{\sqrt{4 k_0^4+k^4_t}} \\
 & = & \ln \left( \frac{\sqrt{4 k_0^4+q^4} + q^2}{2 k_0^2} \right) F(x',q) 
 - \ln \left( \frac{\sqrt{5} + 1}{2} \right) F(x',k_0)
 -   \int_{k_0^2}^{q^2} d k_t^2 \frac{F(x',k_t)}{\sqrt{4 k_0^4+k^4_t}}.
 \label{eq:C}
\eea
Now, at least the $F_C$ term depends {\em only} on the integrated distribution $F(x,k_t)$ that satisfies:
\be
\frac{\partial}{ \partial \ln k^2_t} F(x, k_t) = k^2_t \frac{\partial}{ \partial k^2_t} F(x, k_t) = f (x, k_t)\ .
\ee

Next, we proceed to  study the second term, $F_B$ in (\ref{eq:ABC}), where
\bea
F_B & = & - \int_{k_0^2}^{q^2} \frac{d k_t^2}{k_t^2}
 \int_{k_t^2/z}^\infty d {k'_t}^2
 \frac{k^2_t}{k^{'2}_t}
 \frac{f(x', k_t)}{|{k'_t}^2 - k_t^2|} \\
 &=& \ln (1 - z) \int_{k_0^2}^{q^2} \frac{d k_t^2}{k_t^2} f(x', k_t) \\
  &=& \ln (1 - z) [F(x', q) - F(x', k_0)].
\label{eq:B}
\eea
Here we have a singularity when $z \rightarrow 1$. This singularity will be removed when we include the momentum conservation term (\ref{eq:substr}). It will result in an expression like 
\be
\int \frac{dz}z\ln(1-z)\left[F(x/z,q)-zF(x,q)\right],
\ee
where now the integrand is non-singular at $z\to 1$.

The final contribution, the first term $F_A$ in (\ref{eq:ABC}), is the  most dangerous term of the BFKL kernel. It is given by
\be
\label{eq:uFA}
F_A = \int_{k_0^2}^{q^2} \frac{d k_t^2}{k_t^2}
 \int_{k_0^2}^{k_t^2/z} d {k'_t}^2
 \frac{k^2_t}{k^{'2}_t} \left[\frac{f(x',k'_t)-f(x',k_t)}{|k^{'2}_t-k^2_t|}\right].
\ee
The integral over $k'_t$ of the first term in the above numerator, containing $f(x',k'_t)$, can be evaluated by parts using the identity $ u d v = d( u v) - v du $. In order to see explicitly the cancellation of the singularity at $k'_t=k_t$, we subtract  $F(x,k_t)$ -- a constant independent of $k'_t$ -- from the integrated function. That is, we take
\be
v(x,k'_t)=F(x,k'_t)-F(x,k_t),
\ee
and  integrate the first term in the numerator by parts. Omitting, for the moment, the second variable $x$, for clarity, we obtain
\bea \label{eq:eq6}
F_A = \int_{k_0^2}^{q^2} d k_t^2 \, \left. \left[ \frac{F(k'_t) - F(k_t)}{|{k'_t}^2 - k_t^2|} \right] \right|_{k_0^2}^{k_t^2/z}
  + \int_{k_0^2}^{q^2} d k_t^2 \, \int_{k_0^2}^{k_t^2/z} \frac{d {k'_t}^2}{{k'_t}^2|{k'_t}^2 - k_t^2|} \left[ 
  {k'_t}^2 \frac{F(k'_t) - F(k_t)}{({k'_t}^2 - k_t^2)} - f(k_t) \right].
\eea
Here the first term is well defined when  ${k'_t}^2 = k_t^2$. The second one should be as well, as long as we start with a well defined term in (\ref{eq:uFA}), as is the case. However, we cannot use the linearity property of integration to split the integrand into separate integrals. Instead, we use the Taylor series:
\bea 
F(k'_t) - F(k_t) = \frac{f(k_t)}{k_t^2} ({k'_t}^2 - k_t^2)
 + \left. \left[ \frac{\partial}{\partial {{k'_t}^2}} \frac{f(k'_t)}{{k'_t}^2} \right] \right|_{k'_t= k_t} ({k'_t}^2 - k_t^2)^2
 + \cdots
\eea
to check that the integration is well behaved.
Nevertheless, we still have the unintegrated function $f(k_t)$ in the last term of (\ref{eq:eq6}). However this contribution will disappear after we integrate over $k^2_t$  to obtain finally the integrated distribution. First, we must change the order of integration. We use 
\bea
 \int_{k_0^2}^{q^2} d k_t^2
 \int_{k_0^2}^{k_t^2/z} d {k'_t}^2
 & = &
 \int_{k_0^2}^{q^2/z} d {k'_t}^2
 \int_{z {k'_t}^2}^{q^2} d k_t^2
\eea
and again integrate by parts, this time for $f(k_t)$. We obtain
\begin{align} \label{eq7} \nonumber
  \int d k_t^2 & \frac{1}{|{k'_t}^2 - k_t^2|} \left[ 
  {k'_t}^2 \frac{F(k'_t) - F(k_t) }{({k'_t}^2 - k_t^2)} - f(k_t) \right] 
  =  - k_t^2 \frac{F(k_t) - F(k'_t)}{|{k'_t}^2 - k_t^2|} \\
 & + \int d k_t^2 \, \frac{1}{|{k'_t}^2 - k_t^2|} \left[ 
  {k'_t}^2 \frac{F(k'_t) - F(k_t) }{({k'_t}^2 - k_t^2)} 
  - {k'_t}^2 \frac{F(k_t) - F(k'_t) }{(k_t^2 - {k'_t}^2)} \right] 
\end{align}
The above integrand can be simplified. The result is
\bea \nonumber
 F_A & = & \int_{k_0^2}^{q^2} d k_t^2  \left.\left[ \frac{F(x',k'_t)  - F(x',k_t)}{|{k'_t}^2 - k_t^2|}
\right] \right|^{ {k'_t}^2=k^2_t/z}_{{k'_t}^2=k^2_0}  \\
 & + & \int_{k_0^2}^{q^2/z} d {k'_t}^2  \left. \left[ \frac{k_t^2}{{k'_t}^2} \frac{F(x',k'_t)  - F(x',k_t)}{|{k'_t}^2 - k_t^2|}
\right]\right|^{k_t^2=q^2}_{k_t^2=z{k_t'}^2} .
\eea
To simplify further we apply the limits
\bea \nonumber
F_A & = & \int_{k_0^2}^{q^2} d k_t^2 \left[ \frac{F(k_t/\sqrt{z})  - F(k_t)}{|k^2_t/z - k_t^2|} -\frac{F(k_0)  - F(k_t)}{|k_0^2 - k_t^2|} \right] \\
 & + &  \int_{k_0^2}^{q^2/z}  d {k'_t}^2 \left[  \frac{q^2}{{k'_t}^2} \frac{F(k'_t) - F(q)}{|{k'_t}^2 - q^2|} - z \frac{F(k'_t)  - F(\sqrt{z} k_t')}{|{k'_t}^2 - z{k_t'}^2|}
\right] ,
\eea
and regroup terms
\be
F_A = \int_{k_0^2}^{q^2} d k_t^2 \frac{F(k_t) - F(k_0)}{|k_t^2 - k_0^2|} 
 + \int_{k_0^2}^{q^2/z}  d k_t^2 \frac{q^2}{k_t^2} \frac{F(k_t) - F(q)}{|k_t^2 - q^2|}
 -  \int_{z k_0^2}^{k_0^2}  \frac{d k_t^2}{k_t^2} \frac{F(k_t/\sqrt{z})  - F(k_t)}{|1/z - 1|}.
\label{eq:A}
\ee

The contribution arising from momentum conservation can be rewritten in terms of the integrated gluons $F$ in analogous way.

\section{The final result}
We gather together the results (\ref{eq:A}), (\ref{eq:B}) and (\ref{eq:C}) for $F_A,~F_B$ and $F_C$, and then insert them into (\ref{eq:ABC}).
The result for the BFKL evolution in terms of the integrated gluon density, (\ref{eq:F}), is given by
\bea \nonumber
\label{eq:iBFKLk0}
F(x,q) & = & F(x,k_0) + F_0(x, q) - F_0(x,k_0) \\ \nonumber
 & + & \frac{N_c \alpha_s(q^2)}{\pi} \int_x^1 \frac{d z}{z} \Bigg\{ \int_{k_0^2}^{q^2} d k_t^2 \frac{F(x',k_t) - F(x',k_0)}{|k_t^2 - k_0^2|} 
 + \int_{k_0^2}^{q^2/z}  d k_t^2 \frac{q^2}{k_t^2} \frac{F(x',k_t) - F(x',q)}{|k_t^2 - q^2|}
\\ \nonumber
 & - &  \int_{z k_0^2}^{k_0^2}  \frac{d k_t^2}{k_t^2} \frac{F(x',k_t/\sqrt{z}) - 
 F(x',k_t)}{|1/z - 1|}  +  \ln (1 - z) [F(x', q) - F(x', k_0)] \\ \nonumber
 & + & \ln \left( \frac{\sqrt{4 k_0^4+q^4} + q^2}{2 k_0^2} \right) F(x',q) 
 - \ln \left( \frac{\sqrt{5} + 1}{2} \right) F(x',k_0)
 -   \int_{k_0^2}^{q^2} d k_t^2 \left[ \frac{F(x',k_t)}{\sqrt{4 k_0^4+k^4_t}} \right] \Bigg\} \\
  & - & \text{energy--momentum conservation term}.
\eea
To obtain the energy-momentum conservation term we replace $x'=x/z$ with $x$ in first argument of the distributions $F(x,k_t)$. We use the natural renormalization scale for the QCD coupling $\alpha_s(q^2)$, see e.g.~\cite{CCS}.

The expression (\ref{eq:iBFKLk0}) may be further simplified {\em if} we assume that perturbative QCD evolution can be extrapolated down to $k_0=0$ and $F(x,k_0)$ vanishes at $k_0\to 0$. Then for very small $k_0$
\bea \nonumber
\label{eq:iBFKLstep}
F(x,q) & = & F_0(x, q) + \frac{N_c \alpha_s}{\pi} \int_x^1 \frac{d z}{z} \bigg\{ \int_{k_0^2}^{q^2/z}  \frac{d k_t^2}{k_t^2} \left[  q^2 \frac{F(x', k_t) - F(x', q)}{|{k_t}^2 - q^2|} \right] \\ \nonumber
 & + & \ln (1-z) F(x',q) + \ln \left( \frac{q^2}{k_0^2} \right) F(x',q) \bigg\} \\
  & - & \text{energy--momentum conservation term}.
\eea
We note the apparent $k_0 \to 0$ divergences in the integral, and in  $\ln k_0$ occurring   just before the ``energy--momentum conservation term''. We may join the two divergences together to demonstrate their cancellation. In this way we obtain the relatively simple equation 
\bea \nonumber
\label{eq:iBFKL}
F(x,q) & = & F_0(x, q) + \frac{N_c \alpha_s}{\pi} \int_x^1 \frac{d z}{z} \bigg\{ \int_0^{q^2/z}  \frac{d k_t^2}{k_t^2} \frac{ q^2 F(x', k_t) 
- (q^2 - |q^2 - {k_t}^2|) F(x', q)}{|{k_t}^2 - q^2|} \\
 & + & \ln [ z (1-z)] F(x',q) \bigg\} ~ -~ \text{energy--momentum conservation term}
\eea
Thus we have a BFKL equation for the {\em integrated} gluon distribution, $F$, which
sums up all the leading $(\alpha_s\ln1/x))^n$ contributions. Besides this, the equation includes those next-to-leading terms which provide the energy-momentum conservation during the evolution (or the iterations) and which takes care of the kinematic cutoff (\ref{eq:constraint}). These next-to-leading terms, account for the major part of the NLL  and higher-order corrections to the original BFKL equation.

\section{Discussion and Outlook}
Let us first discuss the equation proposed long ago in~\cite{BF}, since it demonstrates some of the difficulties in obtaining a BFKL equation for an integrated gluon density. Formally eq.(26) of~\cite{BF} should be considered as the $\ln(1/x)$ BFKL evolution equation for the integrated gluon density $F(Q^2,S^2)$ with $S^2=\Lambda^2/x$ and with the splitting kernel $P(Q^2/k^2;\alpha_s(S^2))$ given by eq.(30) of \cite{BF}. This equation is to be compared with our result (\ref{eq:iBFKLk0}) or (\ref{eq:iBFKL}). The first terms, corresponding to the real gluon emission with $k_t<q$, or $\kappa<1$ in~\cite{BF}, are the same. However, there is an important difference due to gluon reggeization, which is not properly accounted for in~\cite{BF} -- reggeization is a higher-twist effect, which  cannot be reproduced by the ``two-particle irreducible diagrams'' considered in~\cite{BF}. Besides this, the kinematical constraint (\ref{eq:constraint}) for real gluon emission is missed in eqs.(26,30) of~\cite{BF}. Moreover, strictly speaking, eq.(26) assumes an infrared cutoff $k_0=0$; that is, it corresponds to the specific limit presented in (\ref{eq:iBFKL}) at the end of previous section. This was not emphasized in~\cite{BF}.

Another problem of eq.(26) is the very strange choice of the argument, $S^2=\Lambda^2/x$, for the QCD coupling $\alpha_s(S^2)$.  Contrary to the natural choice, $\alpha_s(q^2)$ as in (\ref{eq:iBFKLk0}), $S^2$ is very large at low $x$ scales, and has the effect of completely killing the power growth ($x^{-\omega}$) of the BFKL amplitude since the
value of $\omega=\alpha_s\cdot \chi$ decreases as $\alpha_s\propto 1/\ln(S^2)\simeq 1/\ln(1/x)$. So from evolution equation (26), with  
coupling $\alpha_s(S^2)$, we would obtain a cross section which increases as some power of $\ln(1/x)$, but not as a power of $(1/x)$.

Now we give an overview of the structure of unified DGLAP and BFKL evolution. As was emphasized in the introduction, the DGLAP
equation sums up all the leading $(\alpha_s\ln Q^2)^n$ terms, while BFKL accounts for the $(\alpha_s\ln (1/x))^m$ contributions. In general, for many of the interesting processes at the LHC both logarithms are large\footnote{Even for processes which depend directly on the gluon only at moderate and large values of $x$, we need reliable knowledge of the gluon distribution at very small $x$. since the normalisation of the gluon PDF is fixed by
the energy-momentum sum rule. Note that a large part of the total energy of the gluon is hidden in the low $x$ domain where the gluon density is large and grows with decreasing $x$.}, and are of the same order, so we need to consider evolution which takes care of these large logarithms. 
The DGLAP part describes the variation of the Parton Distribution Functions (PDFs) with increasing values of the scale, $Q^2$; while BFKL evolution provides the correct small $x$ behaviour.

Recall that DGLAP evolution starts from parameterised input distributions, PDF$^{\rm input}(x,Q^2_0)$ at a fixed scale $Q^2=Q^2_0$, and includes the trivial boundary condition, PDF=0, at $x=1$. On the other hand, BFKL evolution starts from a gluon distribution at a fixed, but not very small, value $x=x_0$. Its boundary condition, at relatively low values $q^2 =Q^2$, is driven by confinement. Confinement eliminates the gluon at large distances. Therefore, it looks natural for the BFKL case to have an analogous zero boundary condition, $F(x,q=0)=0$, which leads to the simplified form of the evolution given in (\ref{eq:iBFKL})~\footnote{In the more general case, we may use 
(\ref{eq:iBFKLk0}) with some input function $F^{\rm input}(x,q_0)$. In the original BFKL equation for unintegrated gluon density, the `input' function $f_0$ (see (\ref{eq:uBFKL})) reflects the contribution of the lowest-order (Born) diagram. Correspondingly, we do not expect  the integrated input $F^{\rm input}(x,q_0)$ to grow as $x\to 0$. Rather, the growth of the gluon PDF at small $x$ is generated by BFKL dynamics. Thus, anyway, the input distribution will be negligible in comparison with the strongly increasing PDF as we evolve to very low $x$. }. 

Let us accept these, physically motivated, `zero' boundary conditions. Then for unified BFKL-DGLAP evolution, we need DGLAP-like input in a limited interval of $x$ only (say, from $x_0=0.2$ to 1). The BFKL input at fixed $x=x_0$ and $q>q_0$ is obtained now by a straightforward application of the DGLAP equation. The remaining part is the contribution from the small non-perturbative domain of $q<q_0$ and $x>x_0$. Here we may use an extrapolation like
\be
F(x,q<q_0)~=~F(x,q_0)\frac{q^2}{q^2_0\ },
\ee
or, as used in \cite{KMS}, we may introduce a new parameter, $q_a$, to allow for a better matching of the derivative at $q=q_0$
\be
F(x,q<q_0)~=~F(x,q_0)\frac{q^2}{q^2_0}\left(\frac{q^2+q^2_a}{q_0^2+q^2_a}\right)\ .
\ee
  In such an approach the phenomenological input distribution is used to  describe the large $x$ behaviour only, while the low $x$ dependence is {\it completely} generated by BFKL.
 
Following from this development, it looks promising also to consider unified BFKL-DGLAP evolution for the integrated gluon PDF in terms of the variables ($x,\theta$), as  proposed in \cite{E-theta}, instead of the variables $(x,q)$. The coherence of soft gluon emission automatically provides strong-ordering in the opening angle $\theta=q_t/xp$ (where $p$ is the momentum of the incoming proton). On the other hand, both the BFKL and the DGLAP logarithms are actually logarithms coming from an integration over $d\theta/\theta$.  Therefore, for evolution written in terms of $\theta$, we may expect better accuracy already at LO level, with smaller higher-order corrections. Details of such a promising approach are given in \cite{E-theta}, and a recent numerical study of evolution in $\theta$ can be found in \cite{Toton}.

\section*{Acknowledgements}

MGR thanks the IPPP at the University of Durham for hospitality. This work was supported by the Federal Program of the Russian State RSGSS-4801.2012.2.



\thebibliography{}
\bibitem{BFKL} 

V.S. Fadin, E.A. Kuraev and L.N. LipatovPhys. Lett. {\bf B60}, 50 (1975);\\
E.A. Kuraev, L.N. Lipatov and V.S. Fadin, Sov. Phys. JETP {\bf 44}, 443 (1976);\\
E.A. Kuraev, L.N. Lipatov and V.S. Fadin, Sov. Phys. JETP {\bf 45}, 199 (1977);\\
I.I. Balitsky and L.N. Lipatov, Sov. J. Nucl. Phys. {\bf 28}, 822 (1978).
\bibitem{KMS} J. Kwiecinski, A.D. Martin and A. Stasto,  Phys. Rev. {\bf D56}, 3991 (1997).
\bibitem{ABF} G. Altarelli, R. D. Ball and S. Forte, Nucl. Phys. {\bf B742}, 1 (2006)(and the references therein).
\bibitem{BF} R. D. Ball and S. Forte, Phys. Lett. {\bf B405}, 317 (1997).

\bibitem{Ci} M. Ciafaloni, Nucl. Phys. {\bf B296}, 49 (1988).

\bibitem{KMS1} J. Kwiecinski, A.D. Martin and P.J. Sutton, Z. Phys. {\bf C71}, 585 (1996).

\bibitem{Bo} B. Andersson, G. Gustafson and J. Samuelsson, Nucl. Phys. {\bf B467}, 443 (1996).

\bibitem{FLnll} V.S. Fadin and L.N. Lipatov, Phys. Lett. {\bf B429}, 127 (1998).
 
 \bibitem{CWS}  	
Xu-Chang Zheng, Xing-Gang Wu, Sheng-Quan Wang, Jian-Ming Shen, Qiong-Lian Zhang,
 JHEP 1310 (2013) 117.

\bibitem{CCS} M. Ciafaloni, D. Colferai and G.P. Salam, Phys. Rev. {\bf D60}, 114036 (1999);\\
see also, G.P. Salam, Acta Phys. Polonica {\bf B30}, 3679 (1999).


\bibitem{EL} K. Ellis and E. Levin, Nucl. Phys. {\bf B420} (1994) 517.

\bibitem{E-theta} E. de Oliveira, A.D. Martin and M.G. Ryskin,
  arXiv:1404.7670.

\bibitem{Toton} D. Toton, arXiv:1406.0980.
\end{document}